\documentclass[aps,prb,showpacs,floatfix,amsmath,amssymb,
reprint,superscriptaddress]{revtex4-1}
\usepackage{color}
\usepackage{float}
\usepackage{graphicx}
\usepackage{epsfig}
\usepackage{setspace}
\usepackage{amsmath,amssymb}
\usepackage{verbatim}
\usepackage{bibentry}

\def\nnu{{\nonumber}}

\def\nnu{{\nonumber}}
\def\bk{{\mathbf{k}}}
\def\bK{{\mathbf{K}}}
\def\bkt{{\mathbf{\tilde{k}}}}
\def\bR{{\mathbf{R}}}

\begin{document}
\title{Phonon localization in binary alloys with diagonal and off-diagonal disorder: A cluster Green's function approach}
\author{Wasim Raja Mondal}
\affiliation{Jawaharlal Nehru Centre for Advanced Scientific Research, Bangalore 560 064, India.}
\author{ T.\ Berlijn}
\affiliation{Center for Nanophase Materials Sciences, Oak Ridge National Laboratory, Oak Ridge, TN 37831, USA}
\affiliation{Computational Sciences and Engineering Division, Oak Ridge National Laboratory, Oak Ridge, Tennessee 37831, USA}
\author{M.\ Jarrell}
\affiliation{Louisiana State University, Baton Rouge, Louisiana 70803}
\author{N.\ S.\ Vidhyadhiraja} \email{raja@jncasr.ac.in}
\affiliation{Jawaharlal Nehru Centre for Advanced Scientific Research, Bangalore 560 064, India.}

\begin{abstract}
 We report the development and application of a new method for carrying out computational investigations of the effects of mass and force-constant (FC) disorder on phonon spectra. The method is based on the recently developed typical medium dynamical cluster approach (TMDCA), which is a Green's function approach. Excellent quantitative agreement with previous exact diagonalization results establishes the veracity of the method. Application of the method to a model system of binary mass and FC-disordered system leads to several findings. A narrow resonance, significantly below the van Hove singularity, that has been termed as
the boson peak, is seen to emerge for low soft particle concentrations. 
We show, using the typical phonon spectrum, that the states constituting the boson peak  cross over from being completely localized to being extended as a function of increasing soft particle concentration. In general, an interplay of mass and FC disorder is found to be cooperative in nature, enhancing phonon localization over all frequencies. However, for certain range of frequencies, and depending on material parameters, FC disorder can delocalize the states that were localized by mass disorder, and vice-versa. Modeling vacancies as weakly bonded sites with vanishing mass, we find that vacancies, even at very low concentrations, are extremely effective in localizing phonons. Thus, inducing vacancies is proposed as a promising route for efficient thermoelectrics. Finally, we use model parameters corresponding to the alloy system, Ni$_{1-x}$Pt$_x$, and show that mass disorder alone is insufficient to explain the pseudogap in the phonon spectrum; the concomitant presence of FC disorder is necessary.

\end{abstract}

\maketitle


\section{Introduction}

Most forms of disorder in a crystal structure have two essential consequences, namely a randomness in mass, and a concomitant change in the bond strengths. The phonon spectrum is, naturally, affected strongly by the presence of mass and bond disorder, and can exhibit Anderson localization (AL) \cite{PhysRev.109.1492} depending on the nature and strength of disorder, dimensionality and other factors. 
The possibility of phonon localization due to disorder has evoked great interest over the past several decades, and several theoretical and experimental investigations have been carried out.
In recent years, there has been a resurgence of interest in the field due to several direct experimental observations of phonon localization~\cite{PhysRevLett.118.145701,PhysRevLett.113.175501}. The role of AL in the formation of polar nanoregions in ferroelectrics has also generated a lively debate \cite{ferrolocalization}. 
Theoretical investigations of spectral dynamics in mass and force constant (FC)\footnote{We use the term force-constant disorder to imply that the force constants have been randomly chosen and are fixed i.e.\ no attempt has been made to re-compute force-constants based on structural reorganization.} disordered systems received a big impetus with the development of mean-field based approaches \cite{PhysRevB.9.1783,PhysRevB.14.3462,PhysRevB.24.1872,0305-4608-6-6-014}. However, single-site theories are, by construction, incapable of incorporating the full non-local nature of force constants. Additionally, some of the Green's function based attempts failed to maintain Herglotz analytic properties \cite{PhysRevB.18.5291,0022-3719-6-10-003}, which are essential to produce physically acceptable results. Various other extensions of single-site theories\cite{0022-3719-17-6-010,0022-3719-18-22-010,0305-4608-18-10-008,PhysRevB.66.214206} have been attempted. Nevertheless, these perturbative methods are plagued by uncontrollable approximations, and hence are unable to treat AL properly.

As a non-perturbative route to understand AL, exact diagonalization (ED) \cite{PhysRev.175.1201} is the most heavily employed method. Although the method does not suffer from approximations, and yields the disorder averaged spectrum, it does have quite a few disadvantages. The state space increases exponentially with system size implying a severe difficulty in simulating large system sizes. The restriction on system sizes, in turn, leads to difficulties
in obtaining information of AL\cite{PhysRev.175.1201}. 
The consideration of three independent force constants namely $\phi_{AA}, \phi_{AB}$ and $\phi_{BB}$ is necessary for a minimal description of FC disorder in a binary alloy system. Such 
a consideration compounds the computational expense involved in ED calculations. Another important exact method for studying AL of phonons is the transfer matrix method (TMM)\cite{0295-5075-97-1-16007}. Disorder effects in masses and force constants are intertwined with each other in most of the binary alloys, but the state-of-the-art TMM calculations~\cite{0295-5075-97-1-16007} have, thus far, treated
the ionic masses and force constants as uncorrelated variables, which is quite unrealistic. In addition, such an approximation can lead to a violation of sum rules. Thus, despite extensive attempts, a satisfactory, reliable method for studying phonon localization in a strong mass and FC disordered binary alloy is still lacking, which calls for further theoretical development. 

In a recent work\cite{PhysRevB.96.014203}, we described the development and application of a  typical medium dynamical cluster approximation (TMDCA) for investigating the effect of mass disorder on the AL of phonons. In this work, we incorporate the effects of FC disorder into the existing framework, thus taking a step closer to realistic disorder. 
The present study has two objectives, namely (i) the development of a formalism for mass and FC disordered systems, that is non-perturbative, systematically convergent, causal, computationally feasible, and is well-benchmarked; and (ii) the application of this formalism to address several open questions (see below). Our first objective has two parts: to develop (a) the DCA to obtain phonon spectra and (b) the TMDCA 
to investigate phonon localization, in a mass and FC disordered binary alloy. The DCA yields the average
density of states (ADOS), which is experimentally observable and is crucial for a basic understanding of disordered lattice vibrations, at a dramatically less computational cost compared to other methods like ED. Concomitantly, the most salient feature of the TMDCA compared to other theories of localization is its ability in predicting localization based on a single-particle order parameter, namely the typical density of states (TDOS). The main development in this work is a cluster adaptation of the Blackman, Esterling and Berk (BEB) formalism \cite{PhysRevB.4.2412}, which was originally proposed for bond-disordered electronic systems. In this study, we adopt a scalar binary alloy model, that consists of a single branch and a single basis atom within the harmonic approximation. 

In order to assess the validity of the method, we carry out quantitative benchmarks, and find excellent agreement. Subsequently, through an application of the method, we attempt to address the following questions/issues related to the effects of pure FC disorder and the interplay of mass and FC disorder: (1) Structurally disordered glasses are known to exhibit a low frequency anomaly, known as the Boson peak, in the density of states. Such an anomaly has also been observed in disordered lattice models, though the origin of the Boson peak in the two families of systems might be quite different. It has been argued that a purely FC disordered system can also exhibit such behaviour. We carry out a comprehensive analysis of the deviations from low frequency Debye behaviour, and 
ask- What are the reasons and conditions for the emergence of a Boson peak in a binary mass and spring disordered system?   (2) In a related context, are the modes associated with the Boson peak localized or delocalized?  
(3) As shown in our previous work\cite{PhysRevB.96.014203}, lighter isotopic impurities lead to strongly localized, short-wavelength phonons in impurity bands. In this work, we ask whether the effect of FC disorder is to reinforce or negate the localization induced by mass disorder? (4) And finally, as relevant for improving the figure of merit in thermoelectrics, we explore the efficacy of vacancies in realizing strong phonon localization over a broad range of frequencies. 
 
The manuscript is organized as follows. In Section 2, we present our model containing both, the mass and FC disorder, and give a detailed description of the DCA and TMDCA formalism as well as their numerical implementation. In Section 3, we present our results and discussion. We summarize our analysis with some future perspectives in Section 4.

\section{Method}
  
The Hamiltonian for lattice vibrations involving a single basis atom is:
\begin{equation}
H=\sum_{l\alpha}\frac{{p^2_\alpha (l)}}{2M (l)} + \frac {1}{2} \sum_{ll^\prime,\alpha\beta} \Phi^{\alpha\beta}(l,l^\prime)u_\alpha(l)u_\beta(l^\prime)\,,
\label{eq:ham}
\end{equation}
where $u_\alpha(l)$ is the displacement of an ion in the $\alpha^{\rm th}$ direction having mass $M(l)$ in the $l^{th}$ unit cell coupled by the force constant $\Phi_{\alpha\beta}(l,l^\prime)$ tensor to $u_\beta(l^\prime)$.  
We note that the mass $M(l)$ in Eq.\eqref{eq:ham} can vary randomly from site to site. Since we are considering the Hamiltonian in Eq.\eqref{eq:ham} for a binary alloy, the site $l$ can be occupied by either an A-type atom or a B-type atom, {\it i.e} $M(l) \in \{M_A, M_B\}$ with certain probabilities depending on the relative concentrations of the A or B-type atoms. To this end, it is convenient to introduce occupation indices $(x,y)$ for host (A-type atoms) and guest (B-type atoms) as
(following Blackman, Ester and  Berk (BEB)  \cite{PhysRevB.4.2412})

\begin{align}
\begin{split}
x_l=1, y_l=0, & \;\text{if $l \in \rm A $}\\
x_l=0, y_l=1, & \;\text{if $l \in \rm B$}
\end{split}
\end{align}

These occupation indices must obey the following properties:
\begin{align}
\begin{split}
x_{l}y_{l}=0, x_{l}^2=x_l  \\
\langle x_{l}\rangle =c_{\rm A}, \langle y_{l}\rangle =c_{\rm B}
\end{split}
\label{eq:occuprop}
\end{align}
Note that double occupancy of a a given site is prohibited in this formalism.
With this assumption, we are ready to express the randomness in the masses as
\begin{equation}
M(l)=
\begin{cases}
x_l M(l) x_l=M_{\rm A}\\
y_l M(l) y_l=M_{\rm B} \\
x_l M(l) y_l=M_{\rm AB}=0\\
y_l M(l) x_l=M_{\rm BA}=0\\
\end{cases}
\label{eq:diffmass}
\end{equation}

We incorporate such randomness in our formalism by defining a local disorder potential matrix ${\hat{V}}$, as
\begin{equation}
\left({\hat{V}}\right)_{ll^\prime} = \left(1 - M(l)/M_0
\right)\delta_{l,l^\prime}\,,
\label{eq:dispot}
\end{equation}
where $M_0$ is a reference (host) mass, and as a convention, we have chosen A to be the host, hence
$M_0=M_A$.

The corresponding probability distribution for binary disorder reads as
\begin{equation}
P(V_l)=c_{\rm A} \delta(V_l-V_{\rm A})+c_{\rm B}\delta(V_l-V_{\rm B})\,,
\end{equation}
where $c_{\rm A}$ and $c_{\rm B}=1-c_{\rm A}$ are the concentrations of A and B type of atoms  respectively.

Mass disorder can be isotopic or non-isotopic. In general, mass disorder will be accompanied by a {\em corresponding} randomness in the force constants as
\begin{equation}
    \Phi(l,l^\prime)= 
\begin{cases}
   x_l \Phi^{\alpha\beta}(l,l^\prime)x_{l^\prime}=\Phi^{\alpha\beta,\rm AA}(l,l^\prime)\\
    y_l \Phi^{\alpha\beta}(l,l^\prime)y_{l^\prime}=\Phi^{\alpha\beta,\rm BB}(l,l^\prime)\\
    x_l \Phi^{\alpha\beta,}(l,l^\prime)y_{l^\prime}=\Phi^{\alpha\beta,\rm AB}(l,l^\prime)\\
    y_l \Phi^{\alpha\beta}(l,l^\prime)x_{l^\prime}=\Phi^{\alpha\beta,\rm BA}(l,l^\prime)\\
\end{cases}
\label{eq:spring}
\end{equation}

Note that $\Phi(l,l^\prime)$ can be decomposed into diagonal $\Phi(l,l)$ and off-diagonal parts 
\begin{equation}
\Phi^{\alpha\beta}(l,l^\prime) = \delta_{\alpha\beta}(\Phi_D\delta_{l,l^\prime}
 + \Phi_{nn}\delta_{\bR_{l^\prime},\bR_l+\vec{\delta}})\,,
 \label{eq:phiform}
\end{equation}
where $\Phi_D$ and $\Phi_{nn}$ are the diagonal, and the  off-diagonal component of the tensor, respectively, and $\vec{\delta}$ is defined as a vector from a site to its nearest neighbors.
Force constant tensor must obey a sum rule, namely $\sum_{l^\prime} \Phi^{\alpha\beta}(l,l^\prime)=0$.
For satisfying the sum rule, the formalism must incorporate multi-site correlations, because the spring constant tensor is off-diagonal in nature. We are satisfying this property by systematically by increasing $N_c$.

Now, we are in a position to apply the equation of motion method to the Hamiltonian (Eq.\eqref{eq:ham}) and obtain the Dyson equation as
\begin{equation}
M(l) \omega^2 D(l,l^\prime,\omega) = \delta_{ll^\prime}
+\sum_{l^{\prime\prime}l^\prime} {\Phi (l, l^{\prime\prime})} D(l^{\prime\prime}, l^\prime,\omega)\, 
\label{eq:eom}
\end{equation}
Next, we premultiply and postmultiply the above Eq.\eqref{eq:eom} by $x_l$ and $y_l$, which would generate the four possible configurations of the binary alloy. Combining this with Eqs.\eqref{eq:occuprop}, \eqref{eq:diffmass} and \eqref{eq:spring} yields four self-consistent equations for the Green's functions as given below:
\begin{align}
D_{\rm AA}(l,l^\prime) &= \delta_{ll^\prime} 
+ d_{\rm A}(l) \sum_{l^{\prime\prime}\neq l} \Phi^{\rm AA} (l,l^{\prime\prime})D_{\rm AA}(l^{\prime\prime}, l^\prime) \nnu \\
&+ d_{\rm A}(l) \sum_{l^{\prime\prime} \neq l} \Phi^{\rm AB} (l,l^{\prime\prime})D_{\rm BA}(l^{\prime\prime}, l^\prime)\, 
\label{eq:eomAA}\\
D_{\rm AB}(l,l^\prime) &= d_{\rm A}(l) \sum_{l^{\prime\prime}\neq l} \Phi^{\rm AA} (l,l^{\prime\prime})D_{\rm AB}(l^{\prime\prime}, l^\prime) \nnu \\
&+ d_{\rm A}(l) \sum_{l^{\prime\prime} \neq l} \Phi^{\rm AB} (l,l^{\prime\prime})D_{\rm BB}(l^{\prime\prime}, l^\prime)\,
\label{eq:eomAB}\\
D_{\rm BA}(l,l^\prime) &= d_{\rm B}(l) \sum_{l^{\prime\prime}\neq l} \Phi^{\rm BA} (l,l^{\prime\prime})D_{\rm AA}(l^{\prime\prime}, l^\prime) \nnu \\
&+ d_{B}(l) \sum_{l^{\prime\prime} \neq l} \Phi^{\rm BB} (l,l^{\prime\prime})D_{\rm BA}(l^{\prime\prime}, l^\prime)\,
\label{eq:eomBA}\\
D_{\rm BB}(l,l^\prime) &= \delta_{ll^\prime} 
+ d_{\rm B}(l) \sum_{l^{\prime\prime}\neq l} \Phi^{\rm BA} (l,l^{\prime\prime})D_{\rm AB}(l^{\prime\prime}, l^\prime) \nnu \\
&+ d_{\rm B}(l) \sum_{l^{\prime\prime} \neq l} \Phi^{\rm BB} (l,l^{\prime\prime})D_{\rm BB}(l^{\prime\prime}, l^\prime)\,
\label{eq:eomBB}
\end{align}
where the four configuration dependent Green's function are defined as
\begin{align}
& x_l D(l,l^{\prime})x_{l^{\prime}}=D_{\rm AA}(l,l^\prime) \nnu \\
& x_l D(l,l^{\prime})y_{l^{\prime}}=D_{\rm AB}(l,l^\prime) \nnu \\
& y_l D(l,l^{\prime})x_{l^{\prime}}=D_{\rm BA}(l,l^\prime) \nnu \\
& y_l D(l,l^{\prime})y_{l^{\prime}}=D_{\rm BB}(l,l^\prime) \
\end{align}
and the bare locators, $d_{\rm A}$ and $d_{\rm B}$ are given by
\begin{align}
x_l d(l) x_l &=d_{\rm A}(l) =\frac{1}{ \left[M_{\rm A} \omega^2-\Phi^{\rm AA}(l,l) \right]}\\
y_l d(l) y_l &= d_{\rm B}(l) =\frac{1}{\left[ M_{\rm B} \omega^2-\Phi^{\rm BB}(l,l) \right]}
\end{align}

The four self-consistent equations, Eqs.\eqref{eq:eomAA}-Eqs.\eqref{eq:eomBB} may be combined in a convenient $2\times 2$ matrix form as
$$
{
\begin{pmatrix}
D_{\rm AA} & D_{\rm AB} \\
D_{\rm BA} & D_{\rm BB}
\end{pmatrix}
}_{ll^\prime}
=
{
\begin{pmatrix}
d_{\rm A} & 0 \\
0 & d_{\rm B}
\end{pmatrix}
}_{l}\delta_{ll^\prime}\\
+\\
$$
$$
{
\begin{pmatrix}
d_{\rm A} & 0 \\
0 & d_{\rm B}
\end{pmatrix}
}_{l}\\
\sum_{l^{\prime\prime}\neq l}{
\begin{pmatrix}
\Phi^{\rm AA} & \Phi^{\rm AB} \\
\Phi^{\rm BA} & \Phi^{\rm BB}
\end{pmatrix}
}_{ll^{\prime\prime}}
\\
{
\begin{pmatrix}
D_{\rm AA} & D_{\rm AB} \\
D_{\rm BA} & D_{\rm BB}
\end{pmatrix}
}_{l^{\prime\prime} l^{\prime}}
$$
And finally, even this matrix equation can be compactified to get
\begin{align}
\underline{D}=\underline{d}+ \underline{d}\times \underline{\Phi} \times \underline{D} 
\end{align}
Here $\underline{D}$ and $\underline{d}$ are matrices of size $2N\times 2N$, where
$N$ is the system size.

Thus, we have obtained an equation which has a structure similar to the one obtained  in the BEB formalism~\cite{PhysRevB.4.2412} for the electronic problem. It is interesting to note that there is no randomness associated with $\Phi$ matrices. All the randomness is absorbed in the $d$ matrices, and the origin of this randomness lies in the mass term. The $\Phi$ matrices will take the values depending on the random values associated with the mass term. Hence, diagonal mass disorder and off-diagonal spring disorder are dependent on each other. The other point to note is that we can consider three different spring constants, which has been a computational limitation for some theoretical approaches \cite{1742-6596-286-1-012025,PhysRevB.87.134203}. In order to
solve these equations, we have adopted the dynamical cluster approximation. The formalism is similar to the one presented in our previous work on mass disorder, but there are certain steps that are unique to the spring disorder case. In the next section, we provide the details of the formalism.

\subsection{Dynamical Cluster Approximation (DCA)}
The advent of dynamical mean field theory (DMFT) led to a sort of revolution in the understanding of quantum many body lattice
systems. However, since DMFT ignores non-local dynamical correlations, several phenomena such as d-wave superconductivity,
Anderson localization, and low dimensional physics are out of scope of this framework. Hence, quantum cluster approaches, 
that go beyond DMFT, have assumed great importance. One such approach, that is based on momentum space clusters, is the dynamical cluster approximation (DCA).

The DCA may be viewed as an approximation to the wave-vector sums that occur in Feynman-Dyson Perturbation Theory.  Here, the first Brillouin zone containing $N$ wavenumbers $\bk$ is broken into $N_c$ non-overlapping coarse graining cells.  We then approximate the integrals associated with each diagram by its sum of average/coarse-grained estimates of the integrand within each cell.  There is considerable freedom in how this is done.  For example, if the integrand is composed of the product of two functions of the integration variable, do we take the product of the two averages, or the average of the products to define the approximate value in the cell?  Since these two approximations have the same error provided that the number of such cells is large, we can use this freedom to simplify the approximation.  To do this, we define the many-to-few mapping $M(\bk) = \bK$, where $\bK$ labels the cells including $\bk$, so that 
$\bk=\bK+\bkt$, where $\bkt$ labels the wave-numbers within each cell.  The corresponding transformation of the Lie algebra is 
$
{\bar{c}}_\bk = \sum_\bkt c_{\bK+\bkt}\,.
$
It is easy to see that this transformation preserves the Lie algebra.  So it maps bosons onto bosons and fermions onto fermions. The mapping is not canonical though since information is lost in the process.  Nevertheless, this mapping ensures that FDPT may be used to analyze lattice plus the quantum impurity problem.  Under this transformation, all points within each cell are considered to be equivalent, and are mapped to a single point $\bK$.  So, each Green's function within the cell may be replaced by its average value within the cell.  Equivalently, each $G(\bk)$ in a Feynman graph may be replaced by its coarse grained analog.  More significantly, each sum over $\bk$ is replaced by a sum over $\bK$ thereby dramatically reducing the complexity of the problem of order $N$ to order $N_c$.   The associated FDPT is the same as a small self-consistently embedded periodic cluster problem.  Once the cluster problem is solved, we calculate the corresponding irreducible self energy and vertex functions.  We use them in the Dyson and Bethe-Salpeter equation to calculate the single-particle spectra and the two-particle susceptibilities.

For example, the DMFT framework may be represented as a mapping of the entire first Brillouin zone to just one momentum at the centre for the zone. The main simplification in the DMFT  framework is the absence of momentum conservation at the vertices in the Feynman diagrams, thus leading to a local self-energy. The DCA targets this lacuna of DMFT and replaces the Dirac delta function that represents true momentum conservation at the vertices by a Laue function that conserves momentum, but only for the cluster momenta. This brings back the momentum dependence in the Green's functions and self-energy, lost at the DMFT level. Thus, as the number of clusters increases, the Brillouin zone is sampled more
densely, and hence the thermodynamic limit is systematically approached. We refer the reader to review articles \cite{RevModPhys.77.1027,TMDCAreview} for details of the DCA, and its applications to a variety of problems.
The DCA algorithm that we have implemented is derived using the formalism described in the previous subsection and is described below:

1. We start with an initial guess of the hybridization function as
$$
\underline{\Delta (\mathbf K,\omega)}=
\begin{pmatrix}
\Delta^{\rm AA}(\mathbf K,\omega) & \Delta^{\rm AB}(\mathbf K,\omega)\\
\Delta^{\rm BA}(\mathbf K,\omega) & \Delta^{\rm BB}(\mathbf K,\omega)
\end{pmatrix}
$$
This guess may be obtained through a coarse graining of the non-disordered Green's function, or from a previously converged calculation.

2. As a first step for solving the cluster problem, we generate random configurations of the disorder potential V. The disorder potentials $V_{A}$ and $V_{B}$ are assigned depending on whether the site is occupied by A or B type of atom. We generate some random number and if it is less than a given impurity concentration $c_A$, we assign a given site as A-type, else it is assigned as B-type. 

3. We define $\Phi^{\prime}$ as configuration dependent force-constants which can be obtained by configuration dependent Fourier transform as shown below
\begin{equation*}
\Phi^{\prime}(l,l^\prime)= \begin{cases}
        \sum_{\mathbf K} {\big(\underline{\omega^2_{\bK}}\big)^{\rm AA}} e^{i\mathbf K \cdot(\vec R_l-\vec R_{l^\prime})},&\text{if $l \in \rm A$, $l^{\prime}\in \rm A$ }
        \\ 
     \sum_{\mathbf K} {\big(\underline{\omega^2_{\bK}}\big)^{\rm AB}} e^{i\mathbf K \cdot(\vec R_l-\vec R_{l^\prime})},&\text{if $l \in \rm A$, $l^{\prime}\in \rm B$ }.
        \\
         \sum_{\mathbf K} {\big(\underline{\omega^2_{\bK}}\big)^{\rm BA}} e^{i\mathbf K \cdot(\vec R_l-\vec R_{l^\prime})},&\text{if $l \in \rm B$, $l^{\prime}\in \rm A$ }.
       \\
        \sum_{\mathbf K} {\big(\underline{\omega^2_{\bK}}\big)^{\rm BB}} e^{i\mathbf K \cdot(\vec R_l-\vec R_{l^\prime})},&\text{if $l \in \rm B$, $l^{\prime}\in \rm B$ }.
       \end{cases}
 \end{equation*}
where the 2X2 dispersion matrix is given by
$$
{\underline {\omega^2_{\mathbf K}}}=
\begin{pmatrix}
\Phi^{\rm AA}& \Phi^{\rm AB}\\
\Phi^{\rm BA}& \Phi^{\rm BB}
\end{pmatrix}
\bar \omega_{\mathbf K}^2
$$
and the coarse-grained dispersion is given by
\begin{multline*}
\bar \omega_{\mathbf K}^2= \frac{N_c}{N} \sum_{\tilde k}\sin^2 \bigg ( \frac{(K_x +\tilde k_x)a}{2}\bigg)\\
+ \sin^2 \bigg (\frac{(K_y+\tilde k_y)a}{2}\bigg) + \sin^2 \bigg (\frac{(K_z+\tilde k_z)a}{2}\bigg)\\
\end{multline*}
The real space hybridization function $\Delta^{\prime}$ is obtained from the configuration dependent Fourier transform as below:
\begin{equation*}
\Delta^{\prime}(l,l^\prime)= \begin{cases}
        \sum_{\mathbf K}\lbrack \Delta^{\rm AA}(\mathbf K,\omega)\rbrack  e^{i\mathbf K.(R_l-R_{l^\prime})},&\text{if $l \in \rm A$, $l^{\prime}\in \rm A$ }
        \\ 
     \sum_{\mathbf K}\lbrack \Delta^{\rm AB}(\mathbf K,\omega)\rbrack e^{i\mathbf K.(R_l-R_{l^\prime})},&\text{if $l \in \rm A$, $l^{\prime}\in \rm B$ }.
        \\
         \sum_{\mathbf K}\lbrack \Delta^{\rm BA}(\mathbf K,\omega)\rbrack e^{i\mathbf K.(R_l-R_{l^\prime})},&\text{if $l \in \rm B$, $l^{\prime}\in \rm A$ }.
       \\
        \sum_{\mathbf K}\lbrack \Delta^{\rm BB}(\mathbf K,\omega)\rbrack e^{i\mathbf K.(R_l-R_{l^\prime})},&\text{if $l \in \rm B$, $l^{\prime}\in \rm B$ }.
       \end{cases}
 \end{equation*}
After constructing $\Phi^{\prime}$, $\Delta^{\prime}$ and $V$, we compute  the corresponding cluster Green function through the mass-weighted Dyson's equation \cite{PhysRevB.96.014203} as
\begin{align}
& D^c(l,l^\prime,\omega,V)= \nnu \\
& \sqrt{1-(\hat V)_l} (\omega^2 I-\underline{\Phi^{\prime}} -\underline{\Delta^{\prime}} -\underline{\hat V})_{ll^{\prime}}^{-1} \sqrt{1-(\hat V)_{l^\prime}}\,.
\label{eq:mwdysonspring}
\end{align}
\noindent
4. The next step is disorder averaging over disorder configurations denoted by $\langle (...)\rangle$. These
disorder averaged Green's functions correspond to a translationally invariant system, and are denoted by the DCA subscript:
\begin{align}
\big (D^c_{\scriptscriptstyle{\rm DCA}}\big)_{\rm AA} = \bigg\langle  &D^c(l,l^\prime,\omega)  \bigg\rangle, &\text{if $l \in \rm A$, $l\in \rm A$ }\nnu \\
\big (D^c_{\scriptscriptstyle{\rm DCA}}\big)_{\rm AB} = \bigg\langle  &D^c(l,l^\prime,\omega)  \bigg\rangle, &\text{if $l \in \rm A$, $l\in \rm B$ }\nnu \\
\big (D^c_{\scriptscriptstyle{\rm DCA}}\big)_{\rm BA} = \bigg\langle  &D^c(l,l^\prime,\omega)  \bigg\rangle, &\text{if $l \in \rm B$, $l\in \rm A$ }\nnu \\
\big (D^c_{\scriptscriptstyle{\rm DCA}}\big)_{\rm BB} = \bigg\langle  &D^c(l,l^\prime,\omega)  \bigg\rangle, &\text{if $l \in \rm B$, $l\in \rm B$ }\nnu \\
\label{eq:dcaave}
\end{align}
Next, we construct a matrix of the cluster Green function by 
re-expanding the Green function to a $2N_c \times 2N_c$ matrix. It can be represented as
$$
\underline{{D^c_{DCA}}}=
\begin{pmatrix}
\big (D^c_{\scriptscriptstyle{\rm DCA}}\big)_{\rm AA} & \big (D^c_{\scriptscriptstyle{\rm DCA}}\big)_{\rm AB} \\
\\
\big (D^c_{\scriptscriptstyle{\rm DCA}}\big)_{\rm BA} & \big (D^c_{\scriptscriptstyle{\rm DCA}}\big)_{\rm BB}\\
\end{pmatrix}
$$
\noindent
5. As mentioned above, after disorder averaging, the translation symmetry is restored and we can perform Fourier transform for each component to get disorder averaged $\mathbf K$ dependent cluster Green function as
$$
\underline{D^c(\mathbf K, \omega)}=
\begin{pmatrix}
{D^c_{\rm AA}(\mathbf K,\omega) } & { D^c_{\rm AB}(\mathbf K, \omega) } \\
{D^c_{\rm BA}(\mathbf K, \omega)} & { D^c_{\rm BB}(\mathbf K, \omega)}
\end{pmatrix}
$$
\noindent
6. Once the cluster problem is solved, we calculate the coarse-grained lattice Green function as
\begin{align}
& \underline{D^{\rm CG}(\mathbf K, \omega)} \nnu \\
& =\frac{N_c}{N}\sum_{\tilde k}\big \lbrack \underline{D^c(\mathbf K,\omega)}^{-1}+\underline{\Delta(\mathbf K,\omega)}-\underline{\omega^2_k} +\underline{\bar \omega^2 (\mathbf K)}\big \rbrack
\label{coarsegroff}
\end{align}
which in explicit matrix form is given by,
$$
\underline{D^{\rm CG}(\mathbf K, \omega)}
=
\begin{pmatrix}
{D^{\rm CG}_{\rm AA}(\mathbf K,\omega)} & {D^{\rm CG}_{\rm AB}(\mathbf K, \omega)}\\
\\
{D^{\rm CG}_{\rm BA}(\mathbf K, \omega) }& {D^{\rm CG}_{\rm BB}(\mathbf K, \omega)}
\end{pmatrix}
$$
\noindent
7. The DCA self consistency condition requires that the disorder averaged cluster Green function equal the coarse-grained lattice Green's function
\begin{equation}
\underline{D^c(\mathbf K,\omega)}=\underline{D^{\rm CG}(\mathbf K,\omega)}
\end{equation}

\noindent
8. The self consistency condition is used for updating the hybridization function for each component
\begin{align}
&{\Delta_{n}^{\rm AA}}(\mathbf K,\omega)= {\Delta_{o}^{\rm AA}}(\mathbf K,\omega)\nnu \\ &+\xi \left[ {\big (D^c_{\rm AA}(\mathbf K,\omega)\big)}^{-1} - {\big (D^{CG}_{\rm AA}(\mathbf K,\omega)\big)}^{-1}\right]\\
&{\Delta_{n}^{\rm BB}}(\mathbf K,\omega)= {\Delta_{o}^{\rm BB}}(\mathbf K,\omega)\nnu \\
&+\xi \left[ {\big(D^c_{\rm BB}(\mathbf K,\omega) \big)}^{-1}- {\big (D^{CG}_{\rm BB}(\mathbf K,\omega)\big)}^{-1}\right]\\
&{\Delta_{n}^{\rm AB}}(\mathbf K,\omega)= {\Delta_{o}^{\rm AB}}(\mathbf K,\omega)\nnu \\
&+\xi \left[ {\big (D^c_{\rm AB}(\mathbf K,\omega)\big)}^{-1}- {\big (D^{CG}_{\rm AB}(\mathbf K,\omega)\big)}^{-1}\right]\\
&{\Delta_{n}^{\rm BA}}(\mathbf K,\omega)= {\Delta_{o}^{\rm BA}}(\mathbf K,\omega)\nnu \\
&+\xi \left[ {\big (D^c_{\rm BA}(\mathbf K,\omega)\big)}^{-1}- {\big (D^{CG}_{\rm BA}(\mathbf K,\omega)\big)}^{-1}\right]
\label{eq:selfoff}
\end{align}
In the above equations, the $\xi$ is a mixing parameter that determines the fraction of the updated hybridization that should be
mixed with the existing one, thus ensuring smooth convergence of the
DCA iterations. 

As is well known, the Anderson localization of phonons requires us to go beyond DCA. The arithmetic averaging procedure needs to be modified, and a typical averaging ansatz needs to be evolved. Such
an ansatz has been worked out in the electronic case, and has been benchmarked against known results. \cite{PhysRevB.90.094208}.
We have adopted the same ansatz in the phonon case, and have found
that it yields the same level of benchmarks as the electronic case.
The formalism that employs this typical averaging ansatz is called the typical medium DCA or the TMDCA, and is detailed in the next section.

\subsection{Typical Medium Dynamical Cluster Approximation (TMDCA)}
As mentioned in our previous discussion\cite{PhysRevB.96.014203}, the effective medium is constructed via algebraic averaging in the DCA, while the TMDCA utilizes geometric averaging to construct the effective medium. We employ the same ansatz for evaluating the typical density of states as in the electronic case \cite{PhysRevB.90.094208}, as: 
\begin{multline}
\underline{\rho_{typ}^c(\bK, \omega)}= \exp \bigg( \frac{1}{N_c} \sum_{i=1}^{N_c} \langle \rm ln \rho_{ii} (\omega) \rangle\bigg) \times
\\
\begin{pmatrix}
\left\langle \frac{-\frac{1}{\pi} \rm Im D^c_{\rm AA}(\bK,\omega)}{\frac{1}{N_c}\sum_{i=1}^{N_c}\big (-\frac{1}{\pi} \rm Im D^c_{ii} (\omega)\big )}\right\rangle & \left\langle \frac{-\frac{1}{\pi} \rm Im D^c_{\rm AB}(\bK,\omega)}{\frac{1}{N_c}\sum_{i=1}^{N_c}\big (-\frac{1}{\pi} \rm Im D^c_{ii} (\omega)\big )}\right\rangle\\
\\
 \left\langle \frac{-\frac{1}{\pi} \rm Im D^c_{\rm BA}(\bK,\omega)}{\frac{1}{N_c}\sum_{i=1}^{N_c}\big (-\frac{1}{\pi} \rm Im D^c_{ii}(\omega)\big )}\right\rangle & \left\langle \frac{-\frac{1}{\pi} \rm Im D^c_{\rm BB}(\bK,\omega)}{\frac{1}{N_c}\sum_{i=1}^{N_c}\big (-\frac{1}{\pi} \rm Im D^c_{ii}(\omega)\big )}\right\rangle
\end{pmatrix}\,,
\label{eq:anstzoff}
\end{multline}
where $D^c_{ii}$ is defined as
\begin{align}
&D^c_{ii}(\omega)=\sum_{\bf K} \bigg( D^c_{\rm AA}(\mathbf{K},\omega)+D^c_{\rm BB}(\bf K, \omega) \nnu \\
&+D^c_{\rm AB}(\mathbf{K},\omega)+D^c_{\rm BA}(\bf K,\omega) \bigg)\\
 &{\text{ and the local spectral function is given by}} \nnu \\
&\rho_{ii}(\omega)=-\frac{1}{\pi} \left[ D^c_{ii}(\omega)
\right]\,.
\end{align}
Next, we calculate the cluster-averaged typical Green function which is also a $2\times 2$ matrix
\begin{equation}
\underline{{D^c_{\rm typ}}}=
\begin{pmatrix}
\big (D^c_{\scriptscriptstyle{\rm typ}}\big)_{\rm AA} & \big (D^c_{\scriptscriptstyle{\rm typ}}\big)_{\rm AB} \\
\\
\big (D^c_{\scriptscriptstyle{\rm typ}}\big)_{\rm BA} & \big (D^c_{\scriptscriptstyle{\rm typ}}\big)_{\rm BB}\\
\end{pmatrix}
\end{equation}

We compute each component of the cluster-averaged typical Green's function from the corresponding component of the typical density of states \eqref{eq:anstzoff} using the Hilbert transform,
\begin{align}
\big (D^c_{\scriptscriptstyle{\rm typ}}\big)_{\rm AA} = \mathcal{P}\int d\omega^{\prime} \frac{\rho_{typ}^{\rm AA}(\bK,\omega^\prime)}{\omega^2-\omega^{\prime^2}}-i \frac{\pi}{2\omega} \rho_{typ}^{\rm AA}\nnu \\
\big (D^c_{\scriptscriptstyle{\rm typ}}\big)_{\rm AB} = \mathcal{P}\int d\omega^{\prime} \frac{\rho_{typ}^{\rm AB}(\bK,\omega^\prime)}{\omega^2-\omega^{\prime^2}} -i \frac{\pi}{2\omega} \rho_{typ}^{\rm AB}\nnu \\
\big (D^c_{\scriptscriptstyle{\rm typ}}\big)_{\rm BA} = \mathcal{P}\int d\omega^{\prime} \frac{\rho_{typ}^{\rm BA}(\bK,\omega^\prime)}{\omega^2-\omega^{\prime^2}} -i \frac{\pi}{2\omega} \rho_{typ}^{\rm BA}\nnu \\
\big (D^c_{\scriptscriptstyle{\rm typ}}\big)_{\rm BB} = \mathcal{P}\int d\omega^{\prime} \frac{\rho_{typ}^{\rm BB}(\bK,\omega^\prime)}{\omega^2-\omega^{\prime^2}}-i \frac{\pi}{2\omega} \rho_{typ}^{\rm BB}
\label{eq:hilbertoff}
\end{align}
Once the disorder-averaged cluster Green function is calculated using \eqref{eq:hilbertoff}, the self-consistency follows the same steps as in the DCA presented in previous section. The coarse-grained lattice Green's function is then calculated using \eqref{coarsegroff}, which is utilized to update the hybridization function in \eqref{eq:selfoff}.

Using DCA and TMDCA, we can calculate the arithmetically averaged density of states (ADOS) and typical density of states (TDOS) respectively as follows:
\begin{align}
{\rm ADOS}(\omega^2)=-\frac{2\omega}{N_c\pi}{\rm Im} \sum_{{\mathbf K},\sigma\sigma^{\prime}} \big(D^{c}_{\scriptscriptstyle{\rm DCA}}({\mathbf K},\omega)\big)_{\sigma\sigma^\prime} \\
 {\rm TDOS}(\omega^2)=-\frac{2\omega}{N_c\pi}{\rm Im}
 \sum_{{\mathbf K},\sigma\sigma^\prime} \big(D^{c}_{\scriptscriptstyle{\rm typ}}(\mathbf K,\omega)\big)_{\sigma\sigma^\prime}\,,
\end{align}
where $\sigma, \sigma^\prime = A/B$. The formalism described above has been implemented, and we present results and discussion in the following section.

\section{Results and discussions}
\begin{figure}[h!]
    \centerline{\includegraphics[clip=,scale=0.7]
    {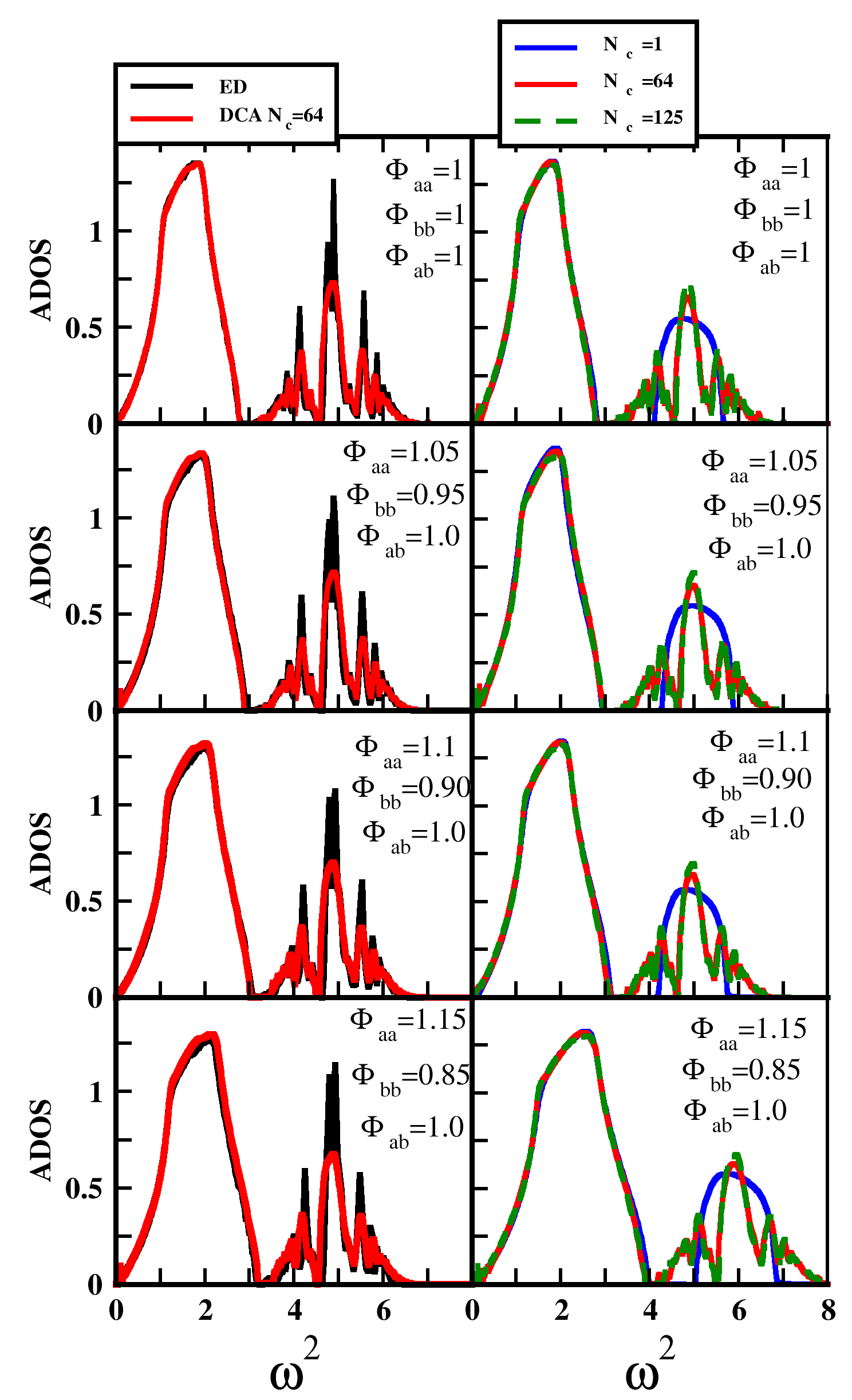}}
\caption{Validation and the convergence of the results using DCA. Left panel:The comparison of the arithmetic density of states (ADOS) obtained from the DCA using N$_c$=64 and ED for various values of $\Phi_{aa}$, $\Phi_{bb}$ and impurity concentration $c$ keeping fixed values of $\Phi_{ab}=\frac{\Phi_{aa}+\Phi_{bb}}{2}$ and disorder potential $V=0.67$ for the three-dimensional binary alloy model. We find a good agreement between the DCA and ED results. Right panel: The evolution of the ADOS for $N_c=1,64,125$ for the same parameter values. Results are converged for $N_c=64$.}
\label{fig:fig1}
\end{figure}
We begin our discussion with a validation of the method. To that end we compute the ADOS via exact diagonalization (ED) of a large number of phonon models of large sized disordered supercells. Within the supercells the impurities are randomly distributed and beyond the supercell boundaries the impurity distributions periodically repeat. Specifically for each set of model parameters we derive the force constant matrices of 100 supercells each with 60 impurities and roughly 400 sites on average. The dynamical matrix of each supercell is then evaluated and diagonalized on a $10\times10\times10$ supercell momentum space grid. We note that not only the impurity distributions but also the shapes of the supercells are randomized under the following constraints. The supercell volumes vary within 375 and 430 sites and the angles between the vectors that span the supercells vary between 75 and 105 degrees. 

The left panels of figure~\ref{fig:fig1} show DCA results (red solid lines) and ED results (black solid lines) for a binary alloy ($c=0.15; V=0.67$) with three different spring constant combinations. A good agreement is seen over all scales thus validating the formalism. We also note that the DCA can access detailed information of the ADOS with relatively small cluster sizes, i.e $N_c=64$ which shows that the DCA is dramatically less expensive compared to ED while being numerically exact. To check the sensitivity of the DCA results on the choice of cluster size $N_c$, we present the ADOS for different $N_c$ in the right panel of Fig.\ref{fig:fig1}. We find that the ADOS for $N_c=64$ and $N_c=125$ are almost identical which implies a rapid convergence of our calculations with respect to increasing $N_c$. In contrast to the results for $N_c=64$ and $N_c=125$, the single-site ($N_c=1$) calculations (solid
blue lines) are unable to capture non-local fluctuations
and disagree significantly with the converged spectra.

The result presented above, namely a comparison of DCA with ED, lends strong credence to results from DCA. Hence, we employ DCA, and subsequently TMDCA to investigate the interplay of spring and mass disorder on phonon spectra and on AL of phonons. We begin with an investigation of the effect of pure spring disorder on phonon spectra.

\begin{figure}[t!]
   \centerline{\includegraphics[clip=,scale=0.7]
    {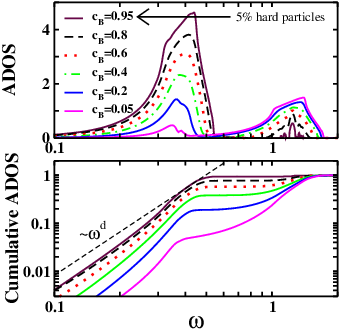}}
\caption{Boson peak appearance at intermediate concentrations -- For pure spring disorder with $V=0.01, \phi_{AA}=1.0, \phi_{BB}=0.1, \phi_{AB}=0.3$, the disorder averaged phonon spectra (top panel) and the corresponding cumulative spectra
(bottom panel) are shown for a range of soft particle concentrations ($c_B=0.05-0.95$). A feature, reminiscent of 
the Boson peak, appears for $c_B\sim 0.2-0.4$.} 
\label{fig:BP_pure_spring}
\end{figure}

The upper panel in figure~\ref{fig:BP_pure_spring}
 shows the average DOS for pure spring disorder (i.e.\ $V\rightarrow 0$) with spring constant values $\phi_{AA}=1.0, \;\phi_{BB}=0.1$ and $\phi_{AB}=0.3$ for various impurity concentrations ($c_B$) ranging from 0.95 to 0.05. The parameters have been chosen to mimic the values obtained in a recent experiment \cite{PhysRevE.87.052301} on crystals of  binary hard-soft microgel particles  with three distinct interparticle potentials. The spring constant values imply that A are hard particles, while B are soft. Hence, $c_B=0.95$
 corresponds to B-particle concentration of 95\% which implies hard sphere concentration of 5\%. As expected, the spectrum for
 a higher concentration of hard particles (stiffer springs, $c_B=0.05$, $c_A=0.95, \phi_{AA}/\phi_{BB}=10$) has almost entire spectral weight at higher frequencies, and as $c_B$ varies from
$0.05$ to $0.95$, spectral weight is transferred to lower frequencies. 
The DOS at 20\% to 40\% soft particles ($c_B=0.2-0.4$) shows a clear excess density of states around a frequency, 
that occurs far below the van Hove singularities of the pure hard particle system. Such behavior is strongly reminiscent of disordered systems, where the origin of such an excess of DOS, termed a Boson peak, has generated a lot of debate. We briefly review a few
theoretical and experimental results relevant to this issue, and place our results in perspective.

It has been shown theoretically \cite{PhysRevLett.81.136}, that a strongly disordered three-dimensional
system of coupled harmonic oscillators with a {\em continuous} force constant distribution exhibits an excess low-frequency DOS (boson peak) as a generic feature. Specifically, if the system
is proximal to the borderline of stability, a low-frequency peak (i.e the Boson peak) appears in the quantity $g(\omega)/\omega^2$ as a precursor of the instability. Our results have been obtained for a binary alloy with three values of spring constants, and we see that a Boson peak appears in a regime of lower soft particle concentration. 

Experimental
measurements \cite{science} of normal modes and the DOS in a disordered colloidal crystal showed Debye-like behavior at low energies and an excess of modes, or Boson peak, at higher energies. The normal modes took the form of plane waves, that hybridized with localized short wavelength features in the Debye regime but lost both longitudinal
and transverse plane-wave character at a common energy near the Boson peak. More recently, experiments\cite{PhysRevE.87.052301} on deformable microgel colloidal particles with random stiffness appear to contradict the theoretical results of Ref\cite{PhysRevLett.81.136}. The authors create crystals of  binary hard-soft microgel particles  with three distinct interparticle potentials distributed randomly on a two-dimensional triangular lattice. The nearest-neighbor bonds are either very stiff ($\phi_{AA}$), very soft ($\phi_{BB}$), or of intermediate stiffness ($\phi_{AB}$). Subsequently, they obtain, experimentally, the phonon modes in crystals with bond strength disorder as a function of increasing dopant concentration. The interesting feature of the microgel crystal is that although the bonds are randomly
distributed, the masses are nearly identical, hence the disorder is purely off-diagonal.  The experimental results\cite{PhysRevE.87.052301} show the absence of a boson peak, although an excess in the density of states as compared to conventional Debye behaviour was observed.

 In the lower panel of figure~\ref{fig:BP_pure_spring}, we present integrated DOS as a function of frequency. The results indicate conventional Debye behaviour ($\sim \omega^d$) at lowest frequencies followed by 
a deviation, and finally a convergence to the normalization value of one at the highest frequencies. In the experiment, the 
hard particle concentration has been varied from 0 to about 21\%. A comparison to figure 3 of Yodh {\it et. al}\cite{PhysRevE.87.052301} shows that 
our results concur well with the experiments. An absence of Boson peak, as concluded in the experiments, is
natural since a clear Boson peak occurs only in the opposite limit of lower soft particle concentration. To summarize,
within the framework of DCA, we find (see figure~\ref{fig:BP_pure_spring}), for a binary alloy, that a transfer of spectral weight to lower frequencies results in the Boson peak, which emerges as a crossover feature
between a pure host system and a pure guest system.

A question that has been much debated in the literature is about the nature of states within the boson peak - Are they localized or delocalized? This question can be effectively answered through the evaluation of the typical DOS, since the typical spectral weight is a measure of the proximity to Anderson localization. A subtle issue about the interpretation of the typical density of states must be mentioned here. A non-zero typical DOS signifies the presence of extended states. According to Mott, a degeneracy of localized and extended states should lead to their hybridization, and hence an eventual delocalization of the localized states. The average DOS and the typical DOS, being different at a given energy, is thus immaterial regarding the identification of the states being extended or localized. 
If the typical DOS is non-zero, the states at that energy should be interpreted as being extended. Concomitantly, a large difference between the average and typical DOS does indicate a proximity to the Anderson localization transition (ALT).

\begin{figure}[t!]
   \centerline{\includegraphics[clip=,scale=0.59]
    {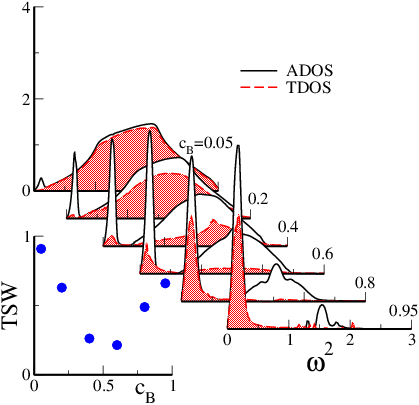}}
\caption{Boson peak -- localized or delocalized? -  For the same parameters as figure~\ref{fig:BP_pure_spring}, the ADOS (solid black line) and TDOS (red shaded part) are shown for concentrations $c_B$ from 0.05 to 0.95. The bottom left panel shows the integrated spectral weight of the typical density of states {\it vs} $c_B$ as solid blue circles.} 
\label{fig:BP_TDOS}
\end{figure}

In figure~\ref{fig:BP_TDOS}, we show, for the same parameters as figure~\ref{fig:BP_pure_spring}, a series of average (black solid lines) and the corresponding typical spectra (red shaded part). 

\begin{figure} [t!]
\centerline{\includegraphics[clip=,scale=0.59]
    {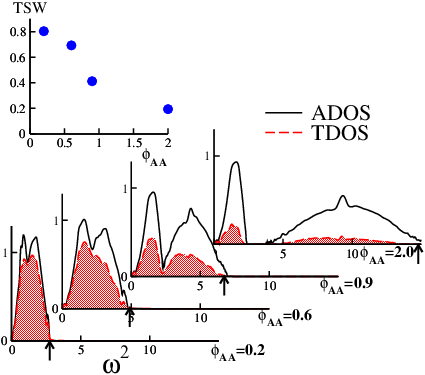}}
\caption{The Average DOS (solid black line) and Typical DOS (red shaded part) for different values of $\phi_{AA}$ is shown. The arrows mark the upper mobility edge. The evolution of the mobility edge and typical spectral weight with increasing values of $\phi_{AA}$ is noticeable: The parameters, $\phi_{BB}=1.0$, $V=0.67$ and $c=0.5$ are fixed, while $\phi_{AB}=0.5(\phi_{AA}+\phi_{BB}$) changes correspondingly. The inset shows the decrease of the integrated typical spectral weight (blue solid circles) with increasing $\phi_{AA}$.}
\label{fig:fig5}
\end{figure}

\begin{figure}[t!]
    \centerline{\includegraphics[clip=,scale=0.57]
    {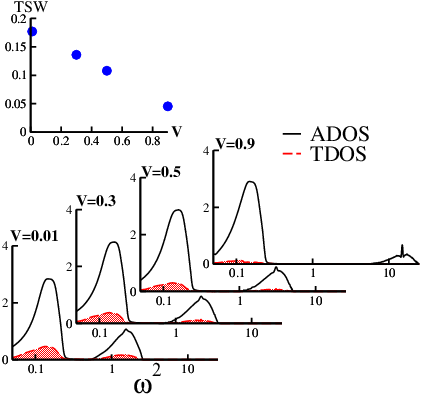}}
\caption{The Average DOS (solid black line) and Typical DOS (red shaded part) for increasing values of $V$ is displayed. The influence of $V$ on the mobility edges and the typical spectral weight is demonstrated: The parameters, $\phi_{AA}=1, \phi_{BB}=0.1, \phi_{AB}=0.3$ and $c=0.5$ are fixed, and the mass ratio of B to A type sites is varied from $\sim$1 to $0.1$, which corresponds to changing $V$ from $0.01$ to $0.9$. The inset shows the decrease of the integrated typical spectral weight (blue solid circles) with increasing $V$.}
\label{fig:fig6}
\end{figure}


The boson peak is seen to have numerically negligible typical spectral weight (TSW) until $c_B\sim 0.4$, beyond which the low frequency peak acquires finite and significant TSW. Thus, the states in the BP exhibit a kind of crossover from being almost localized (proximal to ALT)  to being delocalized (relatively smaller difference between average and typical DOS) with increasing $c_B$. The overall typical spectral weight, shown in the bottom left panel, is non-monotonic, and shows a minimum at $c_B\sim 0.5$, showing that the overall system is closest to the ALT when the ratio of the concentrations of the two species is roughly equal to one.

In order to understand the interplay of mass and spring disorder, we consider two protocols. In the first, we keep the mass ratio parameter, $V=0.67$ and the impurity concentration $c=0.5$ as constants, and increase the spring disorder systematically by varying $\phi_{AA}/\phi_{BB}$
(with $\phi_{BB}=1$, and $\phi_{AB}=\left(\phi_{AA}+\phi_{BB}\right)/2$) from $0.2$ to $2.0$, representing a change of host spring constants from very soft to very stiff. The resulting spectra (ADOS as solid black lines and TDOS as red shaded part) are shown in figure~\ref{fig:fig5}, while the inset  shows the integrated typical spectral weight (TSW, solid blue circles) as a function of $\phi_{AA}$.
For soft A-springs, the characteristic frequencies of the system must be lower than a pure B-type system, and as the $\phi_{AA}$ is increased, spectral weight in the second, high frequency peak increases, as also the bandwidth of the system. So, nominally, it appears that the system is getting delocalized, as the host springs are made stiffer for a fixed mass disorder. However, the inset shows a decrease in TSW with increasing stiffness of $\phi_{AA}$, which implies that the order parameter for AL is decreasing, and hence the system is moving closer to localization.  If we focus on a fixed frequency, say $\omega^2=5.0$, then we see that for $\phi_{AA}=0.2$, the ADOS and TDOS are zero, while for $\phi_{AA}=2.0$, both the average and typical DOS are non-zero, suggesting the interpretation that spring disorder is of a delocalizing nature and counters the localization produced by mass disorder alone. However, the order parameter for AL, namely the TSW, decreases sharply.
Thus the interplay of mass and spring disorder is quite subtle, and an interpretation of the results need to be done carefully. It must be emphasized here, that the subtlety of this interplay has been uncovered through the application of TMDCA, which is able to produce, simultaneously, the typical and the average DOS.

\begin{figure}[t!]
\centerline{\includegraphics[clip=,scale=1.0]
    {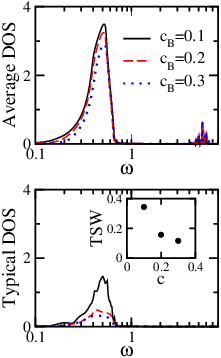}}
\caption{Modeling Vacancies: The spring constants, $\phi_{AA}=1, \phi_{BB}=0.15, \phi_{AB}=0.15$ and the mass ratio, $V=0.95$ are fixed, and three different guest concentrations are considered, namely $c=0.1$(solid black), $c=0.2$(dashed red), and $c=0.3$(dotted blue). The upper panel shows the average DOS, while the lower panel shows the typical DOS. The inset in the lower panel shows the rapid decrease of the integrated typical spectral weight (blue solid circles) with increasing concentration, $c$. }
\label{fig:fig7}
\end{figure}
 

The second protocol is to vary the mass ratio parameter, $V=1-M_{imp}/M_{host}$, keeping the relative concentrations as well the spring constants fixed. The ADOS and TDOS are shown in figure~\ref{fig:fig6} for $V=0.01$ to $0.9$, implying a systematic decrease in the B-site ionic mass. 
Again, lighter impurities imply transfer of spectral weight to higher frequencies, and the B-site band appears as a separate feature, which blue shifts significantly with increasing $V$. In parallel with the results of the first protocol, this result lends itself to an interpretation of delocalization of modes at higher frequencies, but the vanishing of the typical density of states shows that the high frequency modes for $V\rightarrow 1$ are almost localized. The inset shows that the TSW (solid blue circles) decreases sharply with increasing $V$, and this also implies that increasing mass disorder in the presence of fixed spring disorder pushes the system closer to the AL transition. 

The insight we gain from the study of the interplay of mass and spring disorder is that an inference of localization/delocalization of specific modes cannot be made on the basis of ADOS alone, and the TDOS must be concomitantly examined.

 

Vacancies, even at low concentrations, can lead to strong localization of phonons. Within the present theoretical
framework, we model vacancies as weakly bonded sites with vanishing mass. So, we choose $M_{imp}=M_{host}/20$, which is equivalent to $V=0.95$, and spring constants as $\phi_{AA}=1, \phi_{BB}=0.15, \phi_{AB}=0.15$. For these parameters, 
in figure~\ref{fig:fig7}, we show the average DOS (upper panel) and typical DOS (lower panel) for three different guest concentrations, namely $c=0.1$(solid black), $c=0.2$(dashed red), and $c=0.3$(dotted blue). The upper panel shows that the average DOS hardly changes with increasing concentration, while the corresponding typical DOS (lower panel) undergoes significant changes. The inset in the lower panel shows the rapid decrease of the integrated typical spectral weight (blue solid circles) with increasing concentration, $c$. Modeling real vacancies is quite challenging, but the present analysis with a very crude model for vacancies is already indicative of their strong localization effects. The figure of merit for thermoelectrics is inversely proportional to the thermal conductivity, and directly proportional to electrical conductivity. So, in order to maximize the figure of merit, the vacancy concentration, $c$ should be tuned to an optimal value such that it is less than, but not too close to the percolation limit, implying that the electrical conductivity is not too significantly affected, but 
the thermal conductivity due to phonons gets drastically reduced due to the strong localization of acoustic phonons in a large part of the spectrum.  

\begin{figure}[h!]
   \centerline{\includegraphics[clip=,scale=0.6]
    {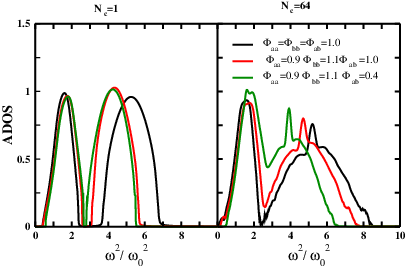}}
\caption{Validation against experiments on NiPt alloy and ICPA. Left panel: ADOS obtained from the DCA method for a three-dimensional binary alloy model at various values of force constants $\Phi_{aa}$, $\Phi_{bb}$ and $\Phi_{ab}$ for fixed impurity concentration $c=0.5$, disorder potential $V=0.67$. Right panel: The corresponding results are shown using same parameter values for $N_c=64$. In the presence of strong spring disorder, the gap in the ADOS obtained from N$_c=64$ reduces which is consistent with  neutron scattering data \cite{PhysRevB.19.2876} as well as previous ICPA results\cite{PhysRevB.66.214206}.} 
\label{fig:DCAalloy}
\end{figure}

Finally, we attempt a qualitative comparison with experiments. It has been argued for a Ni$_x$Pt$_{1-x}$ alloy, that 
 $x=0.65$ constitutes weak force-constant disorder, while
 $x=0.5$ constitutes strong mass and force-constant disorder. Ghosh {\it et. al.}\cite{PhysRevB.66.214206} demonstrate that 
 for $x=0.5$, a CPA-level consideration of inter-atomic force-constants leads to a split-band spectrum. The authors show that a proper treatment of force-constant disorder using
 itinerant CPA leads to a closure of the gap. Our calculations are in full qualitative agreement with these conclusions as argued below. Figure~\ref{fig:DCAalloy} shows 
 ADOS for a binary alloy with M$_{\rm imp}$=M$_{\rm host}$/3
 as appropriate for Ni impurities in Pt host. The left panel
 is for $N_c=1$, equivalent to a CPA calculation, while the
 right panel is for $N_c=64$, which is equivalent to the thermodynamic limit. The impurity concentration used is $c=0.5$, which implies strong mass disorder; and the spectra corresponding to three distinct force-constant combinations are shown. The black solid line corresponds to pure mass-disorder, which at the CPA-level shows a split-band (left panel), while at the DCA-level (right panel), the spectrum has  a two peak  structure with a soft-gap between the peaks. The red and green lines correspond to weak and strong force-constant disorder respectively. Again, the CPA results are
hardly affected by an increase in disorder, while the DCA results for $N_c=64$ show that increasing force-constant disorder leads to significant spectral weight transfer, especially a filling-up of the soft-gap. Since the force constant combination represented by the green line most closely corresponds to the Ni$_{0.5}$Pt$_{0.5}$ alloy, we conclude
that our results agree qualitatively with the ICPA results as 
well as experimental neutron scattering data for Ni$_{0.5}$Pt$_{0.5}$ \cite{PhysRevB.19.2876}.

\section{Conclusions}
The incorporation of the BEB formalism for off-diagonal disorder into the TMDCA yields a reliable, and computationally
feasible approach for investigating binary mass and spring-disordered alloys. Such a conclusion is supported by the benchmarking studies discussed in the initial part of the results section. For a fixed mass ratio, and fixed spring constants ($\phi_{AA}, \phi_{BB}$ and $\phi_{AB}$, increasing the soft particle concentration leads to an excess density of states below the first van Hove singularity of the host hard particle system.
In the present context, it may be identified as the Boson peak, commonly observed in structurally disordered glasses as well as disordered lattice systems, albeit with different origins. We conclude that the origin of the Boson peak in the disordered binary alloy system is a transfer of spectral weight from the guest to the host system, with a necessary condition being the presence of off-diagonal disorder. We emphasize that with pure mass disorder, even though spectral weight transfer does occur, such a BP does not emerge. Additionally, we find that at very low soft particle concentrations, the states in the BP are completely localized, but some of the states crossover to being extended as their concentration is increased. 
The BP eventually ceases to be an anomaly, as the soft particle system becomes the host, and the hard particles assume the role of impurities. The interplay of mass and spring disorder is found to be quite subtle. The overall typical spectral weight decreases upon increasing either of the types of disorder, which indicates that there is a co-operative interplay. However, an added clause is that the interpretation of a co-operative or competitive interplay is also frequency selective, since different parts of the spectrum can transform from being localized to delocalized or vice-versa, depending on the protocol. With a crude modeling of vacancies, we suggest that tuning the concentration of vacancies to an optimal level, which is below but not close to the percolation limit, should be an optimal route to maximizing the figure of merit of thermoelectric materials. All the above results finally culminate in an attempt to understand experiments on Ni$_x$Pt$_{1-x}$ alloys, where, in agreement with ICPA results, we show that $x=0.65$ constitutes weak force-constant disorder, while $x=0.5$ represents a system with strong mass and force-constant disorder. Of course, such a conclusion is qualitative at best, because the real system has a non-trivial structure with multiple branches and spring-constant disorder, while the present implementation is restricted to a scalar approximation and force-constant disorder.
The extension of the present framework to incorporate multiple branches, which will allow us to treat disordered phonons in real materials, is in progress.

\begin{acknowledgments} 
A portion of this research (T.B.) was conducted at the Center for Nanophase Materials Sciences, which is a Department of Energy (DOE) Office of Science User Facility.  This material is also based upon work supported by the National Science Foundation under the NSF EPSCoR Cooperative Agreement No. EPS-1003897 with additional support from the Louisiana Board of Regents (M.J., W.R.M.). N.S.V and W.R.M acknowledge funding from JNCASR, India. M.J acknowledges support from DOE grant DE-SC0017861.
\end{acknowledgments}

\section*{References}
\bibliography{ref}

\end{document}